\documentclass[preprintnumbers,amsmath,amssymb,twocolumn]{revtex4}

\usepackage{amsmath}
\usepackage{amsfonts}
\usepackage{amssymb}
\usepackage{graphicx}

\begin{document}
 
\title{Optimum bias for fast-switching free energy calculations}

\author{Harald Oberhofer}
\author{Christoph Dellago}
\affiliation{Faculty of Physics and Center for Computational Materials Science, University of Vienna Boltzmanngasse 5, 1090 Vienna, Austria}
\date{\today}

\begin{abstract}
We derive the bias function that minimizes the statistical error of free energy differences calculated in work-biased fast-switching simulations. The optimum bias function is compared to other bias functions using a particle pulled through a viscous fluid as an illustrative example. Our analysis indicates that the uncertainty in the free energy is smallest if both dominant and typical work values are sampled with high frequency. 
\end{abstract}

\maketitle



\section{Introduction}

Fast switching computer simulations based on Jarzynski's non-equilibrium work theorem offer an  interesting way for the computation of free energies \cite{Jarzynski1,Jarzynski2,FREE_ENERGY_BOOK}. In this approach, which is particularly relevant in the context of recent mechanical single molecule experiments \cite{Liphardt2002,HummerSzabo2001,Oberhofer2007}, the free energy difference $\Delta F$ between two equilibrium states is related to the work $W$ done on the system during non-equilibrium transformations \cite{Jarzynski1,Jarzynski2}, 
\begin{equation}
\label{eq::biasopt:jar}
e^{-\beta \Delta F}=\left\langle e^{-\beta W}\right\rangle,
\end{equation}
where  $\beta=1/k_{\rm B}T$ is the reciprocal temperature. The angular brackets $\langle \cdots \rangle$ imply an average over many trajectories during which a control parameter is switched at a finite rate between values corresponding to the two equilibrium states. If the control parameter is switched slowly such that the system remains close to equilibrium at all times (in thermodynamics, this corresponds to a reversible transformation), the work done on the system differs little form the free energy difference, $W\approx \Delta F$.  If, on the other hand, the control parameter is switched rapidly, the work values observed in different realizations of the switching process may vary over a large range and typically exceed the free energy difference. In other words, the work distribution $P(W)$ is broad and peaked at work values larger than $\Delta F$ in this case. 

In fast switching computer simulations the exponential average of Equ. (\ref{eq::biasopt:jar}) may create numerical problems particularly in the fast switching regime. The reason for these difficulties is best appreciated if one rewrites the Jarzynski equation (\ref{eq::biasopt:jar}) as integral over the work distribution $P(W)$,
\begin{equation}
\label{eq::biasopt:jar2}
e^{-\beta \Delta F}=\int dW\; P(W) e^{-\beta W}.
\end{equation}
For strong driving, the work distribution $P(W)$ can have a very small overlap with the integrand $P(W) \exp(-\beta W)$ of the above equation. Accordingly, typical work values from the peak of $P(W)$ contribute little to the average while the dominant work contributions to the average are very rare \cite{Jarzynski2006}. This results in large statistical errors of the free energy difference estimated from a finite sample of non-equilibrium trajectories, an issue that has been addressed repeatedly in the recent literature \cite{Oberhofer2007,Jarzynski2006,Lechner2007,Jarzynski2002,Zuckerman2002,Sun1,Ytreberg1,Oberhofer1,Athenes2004,WuKofke2005,Lechner1,Minh2006}. One way to overcome this difficulty consists in favoring the sampling of trajectories with rare but important work values by introducing a work dependent bias function $\Pi(W)$ \cite{Ytreberg1,Oberhofer1}. The bias function guides the simulation, in which fast switching trajectories are harvested using transition path sampling methods \cite{TPS1,TPS2}, toward the important regions of trajectory space. In this approach, Jarzynski's identity takes the form:
\begin{equation}
\label{eq:jarzbias}
e^{-\beta \Delta F}=\frac{\left\langle e^{-\beta W}/\Pi\right\rangle_\Pi}{\langle 1/\Pi\rangle_\Pi},
\end{equation}
where the angular brackets $\langle ... \rangle_\Pi$ denote averages over the biased path ensemble \cite{Ytreberg1,Oberhofer1}. In the present paper, we derive an expression for the bias function that minimizes the statistical error of the free energy estimate. When then test the bias function for a simple one-dimensional model and discuss implications for practical fast switching simulations.


\section{Optimum bias}

In a biased fast switching simulation with bias function $\Pi(W)$ the free energy difference $\Delta F$ is estimated according to Equ. (\ref{eq:jarzbias}) from a finite sample of $N$ trajectories. Accordingly, the free energy estimate $\Delta \overline {F}_N$ is affected by a statistical error quantified by the fluctuations $\epsilon^2_N=\langle(\Delta\overline{F}_N-\Delta F)^2\rangle$, where the angular brackets denote an average over many realizations of the averaging process \cite{Oberhofer1,ZuckermanPRL2002,Bustamante2003}. For large sample sizes $N$ and statistically independent trajectories, the fluctuations $\epsilon^2_N$ are given by \cite{Oberhofer1}
\begin{equation}
\epsilon^2_N=\frac{k_\text{B}^2 T^2}{N}\;\alpha^2 ,
\end{equation}
where the unitless factor 
\begin{equation}
\label{eq::biasopt:alpha}
\alpha^2[\Pi] =	\langle\Pi\rangle \left\langle \frac{\left(e^{-\beta (W -\Delta F)}-1\right)^2}{\Pi}\right\rangle 
\end{equation}
depends only on the work distribution $P(W)$ and the bias function $\Pi(W)$, but not on the sample size $N$. Also, $\alpha^2[\Pi]$ equals the number $N_{kT}$  of trajectories required to obtain a free energy accuracy of $k_{\rm }T$. The argument of $\alpha^2[\Pi]$ emphasizes that the factor $\alpha^2$, which determines the size of the fluctuations for a given sample size $N$, is a functional of the bias function $\Pi(W)$.

We now determine the bias $\Pi^*$ that minimizes $\alpha^2[\Pi]$ and do so by requiring that upon an infinitesimal variation $\delta\Pi$ of the bias function $\Pi^*$ the variation of $\alpha^2$ vanishes,
\begin{equation}
\label{eq::biasopt:alphaconst}
\delta\alpha^2[\Pi^*]=\alpha^2[\Pi^*+\delta\Pi]-\alpha^2[\Pi^*]=0 .
\end{equation}
Expanding the right hand side of Equ. (\ref{eq::biasopt:alpha}) and neglecting all terms of order $(\delta \Pi)^2$ and higher we obtain
\begin{equation}
\delta \alpha^2[\Pi^*] = 
\left\langle\delta\Pi \left[
\frac{\alpha^2[\Pi^*]}{\langle\Pi^*\rangle} -
\frac{(e^{-\beta (W-\Delta F)}-1)^2}{{\Pi^*}^ 2}
\right]\right\rangle .
\end{equation}
Since $\delta \alpha^2[\Pi^*]$ has to vanish for any arbitrary variation $\delta \Pi$, the expression in square brackets must be equal to zero. Solving for $\Pi^*$ then yields 
\begin{equation}
\Pi^*(W)=\frac{\langle\Pi^*\rangle^{1/2}}{\alpha[\Pi^*]} \left\vert e^{-\beta (W-\Delta F)}-1\right\vert,
\end{equation}
where the absolute value $\vert ...\vert$ is taken since the bias function has to be non-negative. Although the fraction on the right hand side is a functional of the (at this point still unknown) optimum bias function $\Pi^*$, it does not explicitly depend on the the work $W$ and can therefore be treated as an irrelevant multiplicative constant. Thus, the optimum bias function can be written as
\begin{equation}
\Pi^*=\left\vert e^{-\beta (W-\Delta F)}-1\right\vert .
\label{equ:optimum_bias}
\end{equation}
This equation is the main result of this paper. Remarkably, the optimum bias function is very general and does not at all depend on the particular switching protocol used in the simulation. It does, however, depend on the unknown free energy difference $\Delta F$ which limits the usefulness of the optimum bias function in practice. For the optimum bias the fluctuations $\alpha^2[\Pi]$ take the particularly simple form
\begin{equation}
\alpha^2[\Pi^*]=\langle |e^{-\beta (W-\Delta F)}-1|\rangle^2.
\end{equation}
Thus, the statistical error in the free energy estimate can be calculated from a single integral over the work distribution.

Due to the non-linearity of the logarithm, the expectation value of the free energy difference estimated from a small sample does not coincide with the true free energy difference. The resulting bias, $b_N=\langle\Delta\overline{F}_N\rangle-\Delta F$, is given by \cite{Oberhofer1}
\begin{equation}
b_N=\frac{k_\text{B}T}{2 N} \langle\Pi\rangle \left\langle \frac{e^{-2 \beta (W-\Delta F)}-1}{\Pi}\right\rangle .
\label{equ:bias}
\end{equation}
As noted earlier  \cite{Lechner2007}, for certain bias functions and work distributions, the bias $b_N$ vanishes. In two situations, this is true also for simulations done with the optimum bias. Consider, first, a switching process for which the work distributions is identical to that of the reverse process. Then, the free energy difference $\Delta F=0$ and it follows from the Crooks theorem \cite{Crooks1} that $\langle (\exp(-2\beta W)-1) /|\exp(-\beta W)-1|\rangle=0$ and hence the bias $b_N$ also vanishes. The other instance in which $b_N$ vanishes concerns processes with Gaussian work distributions, usually observed for slow switching (an exception are isolated systems in which adiabatic invariants prevent the work distribution from becoming Gaussian \cite{Oberhofer1}). In this case, the Jarzynski theorem implies that average work $\overline W$ and work variance $\sigma_W^2$ are related by $\sigma_W^2=2 ( \overline{W}-\Delta F)/\beta$  \cite{Jarzynski1}, and $b_N=0$ can be demonstrated by direct evaluation of the integrals in Equ. (\ref{equ:bias}).


\section{Model}
\label{sec:model}

To illustrate the effect of the optimum bias we applied it to a one dimensional particle dragged through a viscous fluid by a harmonic trap of force constant $k$ translated with constant speed $v$. In this case, the control parameter is the trap position, which changes by $L$ during a time $\tau=L/v$. The particle, whose position is specified by the coordinate $q$, evolves according to the Langevin equation in the overdamped limit\cite{ZWANZIG_BOOK},
\begin{equation}
	\dot{q} = - \frac{k}{\gamma}(q-vt)+\eta,
\end{equation}
where the friction coefficient $\gamma$ is related to the delta-correlated Gaussian random noise $\eta$ by $\langle\eta(0)\eta(t)\rangle=2 k_\text{B}T\gamma^{-1} \delta(t)$. For this model, the work distribution is Gaussian, with mean
\begin{equation}
	\overline{W}=\gamma L v\left[1+\frac{\gamma v}{kL}(e^{-kL/v \gamma}-1)\right] ,
\end{equation}
and variance $\sigma_W^2=2k_{\rm B}T\overline{W}$ \cite{MazonkaJarzynski1999}. All following results were obtained for the parameters $\beta=1$, $\gamma=1$, $k=1$ and $L=5$.


\section{Results}

\begin{figure}[ht]
\begin{center}
	\includegraphics[width=6.5cm,clip=true]{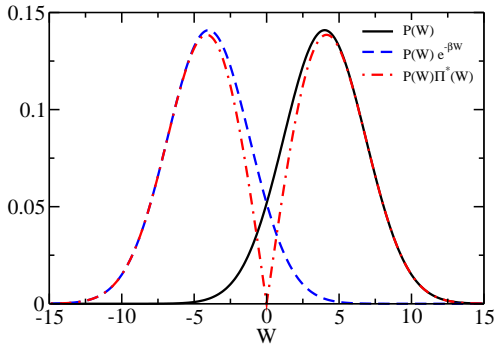}
\end{center}
\caption{Work distribution $P(W)$ (solid line) along with the functions $P(W) e^{-\beta W}$ (dashed line) and $P(W) \Pi^*(W)$ (dash-dotted line) for the particle in the moving harmonic trap. These curves were obtained for the parameter set $\beta=1$, $\gamma=1$, $k=1$ and $L=5$.}
\label{fig::biasopt:bias}
\end{figure}

To visualize the effect of the optimum bias $\Pi^*(W)$ we depict the work distribution $P(W)$ for the particle in the harmonic trap together with $P(W) e^{-\beta W}$, the integrand of Equ. (\ref{eq::biasopt:jar2}), and $P(W) \Pi^*(W)$ in Fig.~\ref{fig::biasopt:bias}. Up to a normalization factor, $P(W) \Pi^*(W)$ is identical to the work distribution $P_{\Pi^*}(W)$ sampled in the biased ensemble,
\begin{equation}
P_{\Pi^*}(W)=P(W) \Pi^*(W) \big/ \int dW\, P(W) \Pi^*(W).
\label{equ:pW_biased}
\end{equation}
For this model, $P_{\Pi^*}(W)$ is symmetric around $W=0$, as follows from the Crooks theorem \cite{Crooks1} for a process with identical work distributions in forward and backward direction. It is interesting to note that according to the work distributions shown in Fig. (\ref{fig::biasopt:bias}), work values from the peaks of $P(W)$ and $P(W)\exp(-\beta W)$ are sampled with the same frequency in the biased ensemble. As discussed in Sec. \ref{sec:conclusion}, this property holds in general and carries ramifications for the design of bias functions.  

We next calculate the statistical error of the free energy estimate as quantified by the factor $\alpha^2$, i.e., the number $N_{kT}$ of trajectories needed to obtain a free energy difference accurate to $k_{\rm B}T$. To estimate the error for different bias functions, we do not carry out actual computer simulations of our model system. Rather, we calculate expected errors from the analytically known work distribution $P(W)$ using Equ. (\ref{eq::biasopt:alpha}). Since in our model the initial and final states correspond to two different position of the otherwise identical trap, the free energy difference vanishes. The expression for the error, Equ.~(\ref{eq::biasopt:alpha}), therefore simplifies to:
\begin{equation}
	\label{eq::biasopt:alphamod}
	\alpha^2[\Pi] = \langle \Pi\rangle \left\langle \frac{(e^{-\beta W}-1)^2}{\Pi}\right\rangle.
\end{equation}
Thus, for given bias function $\Pi(W)$, the error $\alpha^2$ can be easily calculated by integration. 

In addition to the optimum bias derived in this paper, we also examine the exponential bias 
\begin{equation}
\Pi_e(W)=e^{-\beta W/2}
\end{equation}
suggested by Ytreberg and Zuckerman \cite{Ytreberg1}, as well as the inverse bias $\Pi_i(W)$, which flattens the work distribution in the biased ensemble in the interval $[W_{\rm min}, W_{\rm max}]$, 
\begin{equation}
\Pi_i(W)=\left\lbrace
\begin{array}{lll}
1/P(W_\text{min}) & \text{for} & W\leq W_\text{min},\\
1/P(W)& \text{for} & W_\text{min}<W<W_\text{max},\\
1/P(W_\text{max}) & \text{for} & W \geq W_\text{max}.
\end{array}
\right.
\end{equation}
To obtain a flat distribution in the important work range that includes the typical {\em and} dominant work values we choose $W_\text{max,min}=\pm ( \beta \sigma_W^2/2 + 4\sigma_W)$. The case without bias, i.e., $\Pi(W)=1$, is also considered.

\begin{figure}[ht]
\begin{center}
\includegraphics[width=6.5cm,clip=true]{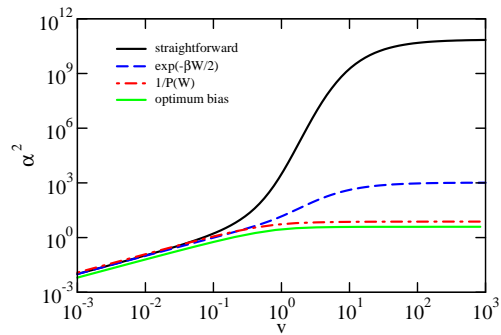}
\end{center}
\caption{The squared fluctuations $\alpha^2$ calculated for the particle in the harmonic trap as a function of the trap velocity $v$ for the various bias functions described in the main text. These curves were obtained for the parameter set $\beta=1$, $\gamma=1$, $k=1$ and $L=5$.}
\label{fig::biasopt:gauss_err}
\end{figure}

For all bias functions discussed here the expected error can be calculated analytically from Equ.~(\ref{eq::biasopt:alphamod}). For the optimum bias one obtains
\begin{equation}
\alpha^2[\Pi^\star]= 4\; {\rm erf}\,^2\left(\frac{\beta \sigma_W}{2\sqrt{2}}\right),
\end{equation}
while for the exponential bias and the inverse bias the error is given by \cite{Lechner1}
\begin{equation}
\label{equ:eps_pi1}
\alpha^2[\Pi_e]=2e^{\beta^2 \sigma_W^2/4}
\left(1-e^{-\beta^2 \sigma_W^2/2}\right),
\end{equation}  
and
\begin{equation}
\label{equ:eps_pi2}
\alpha^2[\Pi_i]=\frac{W_\text{max}-W_\text{min}}{\sigma_W\sqrt{\pi}}
\left(1-e^{-\beta^2 \sigma_W^2/4}\right),
\end{equation}  
respectively. In calculating $\alpha^2[\Pi_i]$ we have assumed that work values outside $[W_{\rm min}, W_{\rm max}]$ do not significantly contribute to the integrals (note that in Ref.~\cite{Lechner2007} the boundaries $W_\text{min}$ and $W_\text{max}$ were selected incorrectly which lead to an erroneous $\alpha^2[\Pi_i]$ for small switching rates). Without any bias the expected error is
\begin{equation}
\label{equ:eps_jar}
\alpha^2[\Pi=1]=e^{\beta^2 \sigma_W^2}-1.
\end{equation}
The resulting values of $\alpha^2$ are depicted in Fig.~\ref{fig::biasopt:gauss_err} as a function of the trap velocity $v$. The statistical error is indeed smallest for the optimum bias $\Pi^*$. While the inverse bias $\Pi_i$ also performs well over the whole range of trap velocities, large errors result for straightforward sampling (no bias) and the exponential bias $\Pi_e$ at large trap velocities. For small trap velocities all bias functions lead to the same asymptotic behavior in which, as for straightforward fast switching without bias, the error depends linearly on the trap velocity as expected from linear response theory. 

\begin{figure}[ht]
\begin{center}
\includegraphics[width=6.5cm,clip=true]{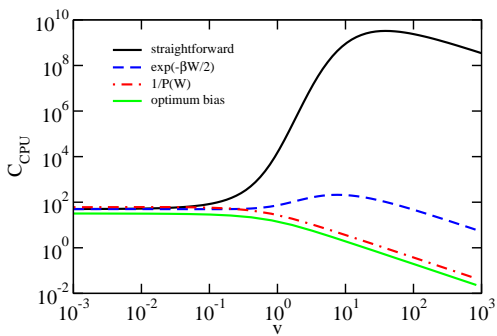}
\end{center}
\caption{Computational cost $C_{\rm CPU}$ calculated for the particle in the harmonic trap with  parameters $\beta=1$, $\gamma=1$, $k=1$ and $L=5$.}
\label{fig::biasopt:ccpu}
\end{figure}

To estimate the total computational cost of the simulation we need to take into account that the cost of a single trajectory is approximately proportional to its length (neglecting any overhead cost, for instance for the generation of initial conditions). Accordingly, we define the computational cost as the product of the number of trajectories $N_{kT}$ and their duration $\tau$ in time \cite{Oberhofer1},
\begin{equation}
C_{\rm CPU}=N_{kT}\, \tau=\alpha^2 \frac{L}{v}. 
\end{equation}
The computational effort $C_{\rm CPU}$ is the total CPU-time required to obtain a free energy accurate to $kT$ in units of the CPU time necessary to compute a trajectory of length 1. As both $N_{\rm kT}$ and $\tau$ depend on the switching rate (i.e., the trap velocity in our case), the benefit of short low-cost trajectories may be compensated by a large number of required trajectories or vice versa. The computational costs resulting from application of the various bias functions are shown in Fig.~\ref{fig::biasopt:ccpu}. For slow switching, the computational cost is constant and very similar for all bias functions implying that from an efficiency point of view it does not matter if one generates a few long trajectories or more but shorter ones (as long as one stays in the linear regime) \cite{Hummer2001}. For larger switching rates, the computational cost declines steadily with the switching rate provided a good bias function is available. This result indicates that the instantaneous switching limit, in which the fast switching method turns into Zwanzig's perturbative approach \cite{Zwanzig1}, yields the most efficient simulations. Note, however, that in this analysis we have neglected correlations, which possibly depend on the switching rate, and any computational overhead involved, for instance, in the generation of initial conditions. Taking such effects into account will abort the decline of $C_{\rm CPU}$ for large switching rates and may well lead to a finite optimum switching rate, but huge savings in computer time are not expected in this case.

\begin{figure}[ht]
\begin{center}
\includegraphics[width=7cm,clip=true]{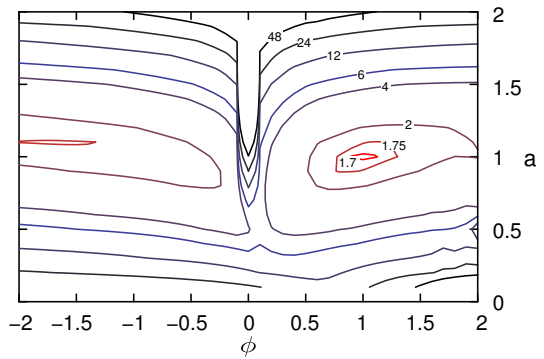}
\end{center}
\caption{Isolines of $\alpha(a, \phi)$ as a function of the parameters $a$ and $\phi$ of the generalized bias $\hat{\Pi}(W)$ with $c=10^{-6}$ for the particle in the harmonic trap moving at velocity of $v=1$.}
\label{fig::biasopt:Esurf}
\end{figure}

Since the optimum bias function depends on the unknown free energy difference (in fact, this is the quantity we wish to calculate), its application is not straightforward. It is therefore interesting to examine the performance of bias functions of similar but more general form.  For this purpose, we introduce the bias function
\begin{equation}
\hat{\Pi}(W)=\left\vert e^{-a \beta W}-\phi\right\vert +c,
\label{equ:generalized}	
\end{equation}
which depends on the parameters $a$, $\phi$ and $c$. In the case of the particle in the harmonic trap, this bias function is identical to the optimum bias function for  $a=1$, $\phi=1$ and $c=0$, while for $a=0.5$, $\phi=0$ and $c=0$ it is equal to the exponential bias. The constant $c$ is added to prevent a singularity in Equ.~(\ref{eq::biasopt:alphamod}) for $\phi>0$. (Note that for the optimum bias no singularity occurs even for $c=0$.) Figure \ref{fig::biasopt:Esurf} shows $\alpha$ as a function of $\phi$ and $a$ for the generalized bias $\hat{\Pi}$ with a trap velocity of $v=1$. The rather shallow minimum of $\alpha$ is indeed located at $\phi=1$ and $a=1$, the values corresponding to the optimum bias. Also for the exponential bias ($\phi=0$, $a=0.5$), the value of $\alpha$ is low, but it is a minimum only on the $\phi=0$ axis. Surprisingly, the $\alpha(\phi,a)$-surface has another minimum at $\phi\approx-1.55$ and $a\approx 1.1$ of almost the same depth as that of the optimum bias function. This is a specific feature of the model and for other models the minimum of $\alpha$ will be located at different values of $\phi$ and $a$.  

\section{Conclusion}
\label{sec:conclusion}

In this article we derived the optimum bias function that leads to the smallest errors in the free energy calculated with work biased path sampling methods \cite{Oberhofer1}. For a one-dimensional model, we compared the accuracy obtained with various bias functions functions confirming that the optimum bias leads to the smallest statistical errors. Remarkably, the optimum bias function is of a very  general form that is independent of the particular switching process (in contrast, the inverse bias $\Pi_i(W)$ is strongly model dependent). The optimum bias does, however depend on the free energy difference, i.e., the very quantity one wants to calculate. This limits the practical applicability of the optimum bias. Nevertheless, some general conclusions can be drawn from the form of the optimum bias function. 

According to Equs. (\ref{equ:optimum_bias}) and (\ref{equ:pW_biased}), for the optimum bias function the work distribution in the biased ensemble is given by
\begin{equation}
P_{\Pi^\star}(W)\propto \left\vert P_R(-W)-P(W)\right\vert,
\label{equ:Pbias}
\end{equation}
where we have used the Crooks theorem \cite{Crooks1}, $P_R(-W)=P(W)e^{-\beta (W-\Delta F)}$,  to relate the work distribution $P(W)$ to that of the switching process carried out with time reversed switching protocol, $P_R(W)$. It has been noted before \cite{Jarzynski2006,Ritort2004}, that $P_R(-W)$ is peaked at the dominant work values, i.e, those work values that mostly contribute to the exponential work average of Equ. (\ref{eq::biasopt:jar2}). Thus, it follows from Equ. (\ref{equ:Pbias}) that both the dominant {\em and} the typical work values are sampled with the same frequency for the optimum bias. This is particularly apparent if the peaks of $P(W)$ and $P_R(-W)$ are far apart and the overlap between the two distributions is small. Then, $P_{\Pi^\star}(W)\approx P_R(-W)$ around the dominant work values and $P_{\Pi^\star}(W)\approx P(W)$ in the range of typical work values. In this case, $\alpha^2[\Pi^*]=(\int dW | P_B(-W)-P(W)|)^2 \approx 4$, such that the number $N_kT$ of required trajectories converges to a constant value in the fast switching limit provided the optimum bias is used. Equation (\ref{equ:Pbias}) also implies that the work values between the dominan++t and typical ones are relatively unimportant. In particular, the work value $W=\Delta F$ does not need to be sampled at all. 

From this analysis of the optimum bias function one may infer that the most important property of good bias functions is that they lead to a biased ensemble in which both dominant and typical trajectories occur with high frequency. At first sight it seems easy to devise a procedure that does exactly that even without explicitly using as bias function: just run an equal number of trajectories in forward and backward direction starting from the respective equilibrium initial conditions. Implicitly, this procedure corresponds to a bias function $\Pi(W)=\exp\{-\beta(W-\Delta F)\}+1$ and to the parameters $a=1$, $\phi =-1$, and $c=0$ of the generalized bias function $\hat \Pi(W)$. As can be seen in Fig. \ref{fig::biasopt:Esurf}, $\alpha(a, \phi)$ is very low but not a minimum at these parameter values since for slightly different parameters ($a=1.1$, $\phi =-1.55$) dominant and typical work values are sampled equally well, but work values in between are sampled less. Although it is easy to generate work values according to $\Pi(W)=\exp\{-\beta(W-\Delta F)\}+1$ using the procedure lined out above (without knowing the bias function itself), problems appear somewhere else in this case. According to Equ. (\ref{eq:jarzbias}) each contribution in the biased ensemble must be corrected by division through the bias function. Since the bias function itself depends on the unknown free energy difference $\Delta F$, however, the correction cannot be carried out.

The above analysis makes also clear why the inverse bias, $\Pi_i(W)=1/P(W)$, works well: if all work values in a sufficiently large range are sampled with the same frequency, the dominant and typical work values occur with approximately the same weight as required. The work values in between are sampled more than needed, but that does not strongly affect the error. For a Gaussian work distribution, the exponential bias $\Pi_e(W)=\exp(-\beta W/2)$ leads to a preferred sampling of the work values between the typical and dominant ones. For slow driving both, typical and dominant work values are included, but for strong driving neither ones are sampled with sufficient frequency leading to substantial statistical uncertainties. Also the generalized bias $\hat \Pi(W)$ of Equ. (\ref{equ:generalized}) leads to small errors if the resulting work distribution in the biased ensemble spans both the dominant and typical work values. The specific parameters $\phi$ and $a$ at which that happens are, however, strongly model dependent such that no generally valid bias function can be deduced. In conclusion, to reduce statistical errors bias functions for fast switching simulations must be designed such that both dominant and typical work values are sampled with high frequency. Devising such bias functions without knowledge of the free energy is challenging.


\section*{Acknowledgements}
This work was supported by the Austrian Science Fund (FWF) under grant No. P17178-N02 and within the Science College "Computational Materials Science" under grant W004. 

\bibliography{bibliography}
\end{document}